\documentclass[aps,pra,preprint,groupedaddress,amsmath,amssymb,showpacs]{revtex4-1}
\usepackage{amssymb}
\usepackage{booktabs}
\usepackage{graphicx,amsmath,amssymb,epsfig,color}
\usepackage{dcolumn}
\usepackage{bm}
\begin{document}
\title{Storage and Retrieval of Light Pulses in a Fast-Light Medium via Active Raman Gain}
\author{Datang Xu$^{1}$, Zhengyang Bai$^{1}$,  and Guoxiang Huang$^{1,2, }$\footnote{gxhuang@phy.ecnu.edu.cn} }
\affiliation{$^1$State Key Laboratory of Precision Spectroscopy, School of Physical and Material Sciences,
                 East China Normal University, Shanghai 200062, China\\
             $^2$NYU-ECNU Joint Institute of Physics at NYU-Shanghai, Shanghai 200062, China
             }
\date{\today}

\begin{abstract}

We propose a scheme to realize the storage and retrieval of light pulses in a fast-light medium via a mechanism of active Raman gain (ARG). The system under consideration is a four-level atomic gas interacting with three (pump, signal and control) laser fields. We show that a stable propagation of signal light pulses with superluminal velocity (i.e. fast-light pulses) is possible in such system through the ARG contributed by the pump field and the quantum interference effect induced by the control field. We further show that a robust storage and retrieval of light pulses in such fast-light medium can be implemented by switching on and off the pump and the control fields simultaneously. The results reported here may have potential applications for light information processing and transmission using fast-light media.

\end{abstract}
\maketitle


\section{Introduction}{\label{Sec1}}

In past two decades, much attention has been paid to the study of slow light via electromagnetically induced transparency (EIT), a typical quantum interference effect occurring in resonant multi-level atomic systems~\cite{Flei}. EIT and slow light have considerable applications,
one of which is optical quantum memory~\cite{Sim,Lvo,San} (defined as a system capable of storing a useful quantum state via interaction with light), very promising for quantum repeaters, quantum computations, and quantum metrology, etc~\cite{Bus}. The optical quantum memory protocol through EIT-based slow-light media utilizes the significant dispersion variation in the transparency window induced by a control field to modify the group velocity, and stop, store, and retrieve a signal light pulse in a controllable way~\cite{Fleischhauer2000,Liu2001,Eisaman2005,Gorshkov2007,Novikova2012,Chenyang}.

Parallel to the study of slow light via EIT, in recent years there has also been great interest on the study of fast light (also called superluminal light)~\cite{note00} via active Raman gain (ARG) or similar mechanisms~\cite{Garrett1970,Steinberg1993,Bolda1994,Wang2000,Stenner2003,Boyd2002,Milo}. Different from EIT-based slow light scheme, where the signal field operates in an absorption mode and hence its attenuation during propagation can not be avoided, in ARG-based fast light scheme the signal field operates in a stimulated Raman emission mode. It is this emission mode that leads to many attractive features~\cite{Deng1}, including non-attenuated propagation, superluminal group velocity, and large, rapidly responding Kerr nonlinearity of the signal field, which can be used to design fast quantum phase gates and form superluminal optical solitons, etc~\cite{Deng1,Aga,Jiang,Huang,Deng2}. A natural question arises: Is it possible to use fast-light media~\cite{note000} to realize light quantum memory?

In a series of interesting works, Akulshin {\it et al.}~\cite{Akulshin2005,Lezama2006,Akulshin2010} showed, both experimentally and theoretically, that light memory in a fast-light medium is indeed possible, and such light memory can be understood as a consequence of transient response of coherently prepared atomic system to optical excitation conditions, and hence light memory is not necessarily restricted to slow-light media but can also occur in fast-light media. However, the light memory scheme in the fast-light medium used in Refs.~\cite{Akulshin2005,Lezama2006,Akulshin2010} is based on electromagnetically induced absorption (EIA), in which the signal light pulse undergoes a significant absorption during propagation and hence the efficiency of the light memory obtained in Refs.~\cite{Akulshin2005,Lezama2006,Akulshin2010} is low. For practical applications, it is desirable not only to obtain a stable propagation of fast-light pulses, but also to realize robust storage and retrieval of light pulses through fast-light media.

In this work, we suggest a new scheme to realize storage and retrieval of light pulses in a fast-light medium. The system we consider is a coherent four-level atomic gas interacting with three (pump, signal, and control) laser fields.  We show that in such a system the gain of the signal field can be well controlled to a very small value by the quantum interference effect induced by the control field, and hence a stable propagation of signal light pulses with a superluminal velocity can be obtained through the ARG contributed by the pump field. We also show that robust storage and retrieval of the light signal pulses in such fast-light medium can be achieved by switching on and off the pump and the control fields {\it simultaneously}.

Our ARG scheme for light memory has the following characteristics: (i)~It provides a new mechanism for light memory, fundamentally different from those based on EIT, CRIB (controlled reversible inhomogeneous broadening)~\cite{Alexander2006,Kraus2006,Staudt2007}, AFCs (atomic frequency combs)~\cite{Riedmatten2008,Afzelius2009}, and others~\cite{Zhu2007,Camacho2009,Hosseini2009};
(ii)~It is also much different from the light memory scheme using the fast-light medium based on EIA reported in Refs.~\cite{Akulshin2005,Lezama2006,Akulshin2010}. In particular, the serious optical absorption inherent in the EIA scheme is completely avoided, and hence the memory efficiency in our ARG scheme is much higher than that in the EIA scheme.
(iii)~It is nearly free of noise due to four-wave mixing, which is usually inevitable in the EIT scheme~\cite{Lauk2013} (and the associated Raman scheme~\cite{Kozhekin2000,Nunnet2007,Sprague2014,Saun}); (iv)~It can work at room temperature and is immune from Doppler effect and the spontaneous emission from excited state;
(v)~Comparing with EIT and other schemes~\cite{Fleischhauer2000,Liu2001,Eisaman2005,Gorshkov2007,Novikova2012,Chenyang,Akulshin2005,Lezama2006,Akulshin2010,Alexander2006,Kraus2006,Staudt2007,Riedmatten2008,Afzelius2009,
Zhu2007,Camacho2009,Hosseini2009,Lauk2013,Kozhekin2000,Nunnet2007,Sprague2014,Saun}, our ARG scheme has more adjustable system parameters because there are two (i.e. the pump and control) fields that can be manipulated simultaneously, which can be used not only to protect the signal field from attenuation and deformation, but also to increase the efficiency and fidelity of the light memory; (vi)~Different from EIT-like schemes,
our ARG scheme has shorter system response time (see Deng {\it et al.}~\cite{Deng1} and Li {\it et al.}~\cite{Deng2}), hence the ARG scheme for light memory may have potential applications for rapidly responding quantum information processing and transmission.

The remainder of the paper is arranged as follows. Sec.~\ref{Sec2} gives a description of the model under study. Sec.~\ref{Sec3} presents the result on stable propagation of fast-light pulses in the system and investigates the storage and retrieval of light pulses using the fast-light medium. A theoretical explanation of these results is also illustrated. Finally, Sec.~\ref{Sec4} is a discussion and summary of our work.

\section{Model}{\label{Sec2}}

\begin{figure}
\centering
\includegraphics[scale=0.7]{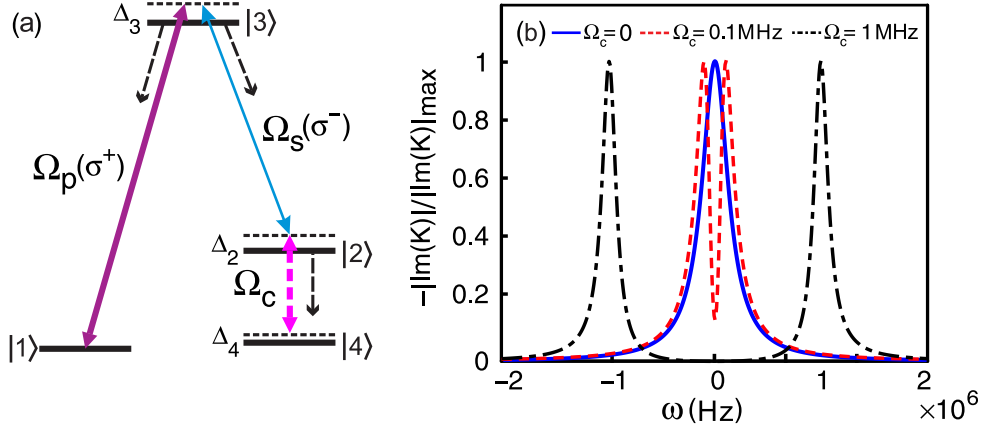}
\caption{(Color online) (a) Energy-level diagram and excitation scheme of the four-level system with ARG. $\Omega_{\rm p}$, $\Omega_{\rm s}$, and $\Omega_{\rm c}$ are respectively half Rabi frequencies of the pump ($\sigma^+$-polarization), signal ($\sigma^-$-polarization), and control
(microwave) fields, which couple respectively the states $|1\rangle$ and $|3\rangle$, $|2\rangle$ and $|3\rangle$, and $|2\rangle$ and $|4\rangle$. $\Delta_3$, $\Delta_2$, and $\Delta_4$ are one-, two-, and three-photon detunings, respectively. Dashed arrows represent the spontaneous emission or incoherent population exchange rates. (b) Gain spectrum of the signal field, $-\text{Im}(K)/|\text{Im}(K)|_{\rm max}$, as a function of
$\omega$~\cite{note01}. The solid (blue), dashed (red), and dashed-dotted (black) lines are for $\Omega_{\rm c}=0$,  0.1~MHz, and 1~MHz,
respectively.}\label{fig1}
\end{figure}

We consider a lifetime-broadened atomic gas with the level
configuration shown in Fig.~\ref{fig1}(a), in which
%
%
a pump laser field, which has a $\sigma^+$-polarization and
the center angular frequency (wavenumber) $\omega_{\rm p}$\,\, ($k_{\rm p}$),
couples the quantum states $|1\rangle$ and $|3\rangle$, a weak
signal laser field, which has a $\sigma^-$-polarization and the
center angular frequency (wavenumber) $\omega_{\rm s}$\,\, ($k_{\rm s}$),
couples the quantum states $|2\rangle$ and $|3\rangle$, and a strong
(microwave) control field,  which has the center angular frequency
(wavenumber) $\omega_{\rm c}$\,\, ($k_{\rm c}$), couples the quantum states
$|2\rangle$ and $|4\rangle$, respectively. $\Omega_{\rm p}$, $\Omega_{\rm s}$,
and $\Omega_{\rm c}$ are respectively half Rabi frequencies of the pump,
signal, and control fields; $\Delta_3\equiv\omega_{\rm p}-(\omega_3-\omega_1)$,
$\Delta_2\equiv\omega_{\rm p}-\omega_{\rm s}-(\omega_2-\omega_1)$, and
$\Delta_4\equiv\omega_{\rm p}-\omega_{\rm s}-\omega_{\rm c}-(\omega_4-\omega_1)$
are respectively the one-, two-, and three-photon detunings, with $\hbar \omega_j$ the
eigenenergy of the state $|j\rangle$ ($j=1,2,3,4$); Dashed
arrows represent the spontaneous emission or incoherent population
exchange rates. The states $|1\rangle$, $|2\rangle$, $|3\rangle$,
and the pump field ($\Omega_{\rm p}$) and the signal fields ($\Omega_{\rm s}$) constitute a
typical ARG core~\cite{Deng1,Jiang,Huang}.

The total electric field in the system is $\mathbf{E}={\bf E}_{\rm p}+{\bf E}_{\rm s}+{\bf
E}_{\rm c}=\sum_{l=\text{p,s,c}}\mathbf{e}_l{\cal E}_l {\rm exp}[i( k_l
z-\omega_lt)]+$c.c., where $\mathbf{e}_l$ is the unit polarization
vector of the electric field ${\bf E}_l$, which has the envelope ${\cal
E}_l$  $ (l=\text{p, s, c})$.  For suppressing Doppler effect, the incident direction of all
three laser fields have been assumed to be along the $z$ direction.
Under electric-dipole and rotating-wave approximations, the Hamiltonian of
the system in interaction picture reads
\begin{equation}\label{Hamiltonian}
\hat{\cal H}_{\rm int}=-\hbar\left[\sum_{j=1}^4 \Delta_j|j\rangle\langle
j|+(\Omega_{\rm p}|3\rangle\langle1|+\Omega_{\rm s}|3\rangle\langle2|+\Omega_{\rm c}|4\rangle\langle2|+{\rm
h.c.})\right],
\end{equation}
where h.c. represents Hermitian conjugation, three half Rabi
frequencies are respectively defined by $\Omega_{\rm p}=(\boldsymbol{p}_{31}\cdot {\bf e}_{\rm p}){\cal
E}_{\rm p}/\hbar$, $\Omega_{\rm s}=(\boldsymbol{p}_{32}\cdot {\bf e}_{\rm s}){\cal
E}_{\rm s}/\hbar$, and $\Omega_{\rm c}=(\boldsymbol{p}_{42}\cdot {\bf e}_{\rm c}) {\cal
E}_{\rm c}/\hbar$, with $\boldsymbol{p}_{jl}$ the electric-dipole
matrix element associated with the transition from the state
$|j\rangle$ to the state $|l\rangle$.  The dynamics of the atoms in the system is
governed by the optical Bloch equation~\cite{Boyd,Zhu}
\begin{equation}\label{Blochequation}
i\hbar\left(\frac{\partial}{\partial
\tau}+\Gamma\right)\sigma=[\hat{\cal H}_{\rm int},\sigma],
\end{equation}
where $\tau=t-z/c$ is traveling coordinate with $c$ the light speed
in vacuum, $\sigma$ is a $4\times4$ density matrix, and $\Gamma$ is
a $4\times4$ relaxation matrix describing the spontaneous emission
and dephasing. The explicit expression of Eq.~(\ref{Blochequation})
is presented in Appendix A.

The propagation of the signal field envelope is governed by the Maxwell
equation, which under the slowly varying envelope approximation
obeys~\cite{Boyd,Zhu}
\begin{equation}\label{Maxwell}
i\frac{\partial \Omega_{\rm s}(z,\tau)}{\partial
z}+\kappa_{23}\,\sigma_{32}(z,\tau)=0,
\end{equation}
where $\kappa_{23}={\cal{N}}_a\omega_{\rm p}|\mathbf{p}_{23}|^2/(2\varepsilon_0c\hbar)$, with
${\cal{N}}_a$ the atomic density. Note that in deriving the above equation the signal-field envelope is assumed to be wide enough in the transverse (i.e. $x$, $y$) directions or the atomic gas is filled within a quasi-one-dimensional waveguide, so that the diffraction effect for the light propagation can be omitted.

\section{Storage and retrieval of fast light pulses}{\label{Sec3}}

\subsection{Stable propagation of fast light pulses}{\label{Sec31}}

Since the signal field is weak (but has enough photon number so that a semi-classical description
is allowed), we can solve the Maxwell-Bloch (MB) equations (\ref{Blochequation}) and (\ref{Maxwell}) by using a  linear approximation.
For simplicity, we assume the one-photon detuning $\Delta_3$ is much larger than the all Rabi frequencies, Doppler-broadened linewidth, atomic coherence decay rates, and frequency shift induced by the pump and the control fields, so the Doppler effect even when the system works at room
temperature can be largely suppressed~\cite{Deng1,Jiang,Huang}.

To gain a foresight for the propagation of the signal field, we must know firstly
the steady-state solution of the system before the signal field is applied, which is easily
obtained by solving the MB Eqs.~(\ref{Blochequation}) and (\ref{Maxwell}) for the case of $\partial/\partial t=0$ and $\Omega_{\rm s}=0$. The result reads
$\sigma_{11}^{(0)}=\gamma_{14}|\Omega_{\rm c}|^2(\Gamma_3X_{31}+|\Omega_{\rm p}|^2)/D$,
$\sigma_{22}^{(0)}=\Gamma_{23}|\Omega_{\rm p}|^2 |\Omega_{\rm c}|^2/D$,
$\sigma_{33}^{(0)}=\gamma_{14}|\Omega_{\rm c}|^2|\Omega_{\rm p}|^2/D$,
$\sigma_{44}^{(0)}=\Gamma_{23}|\Omega_{\rm c}|^2|\Omega_{\rm p}|^2/D$,
$\sigma_{31}^{(0)}=-\Omega_{\rm p}\gamma_{14}\Gamma_3X_{31}|\Omega_{\rm c}|^2/(d_{31}D)$,
$\sigma_{42}^{(0)}=0$,
with $X_{31}=|d_{31}|^2/(2\gamma_{13})$,
$X_{42}=|d_{42}|^2/(2\gamma_{42})$, and
$D=\gamma_{14}|\Omega_{\rm c}|^2(\Gamma_3X_{31}+2|\Omega_{\rm p}|^2)+2\Gamma_{23}
|\Omega_{\rm p}|^2|\Omega_{\rm c}|^2$. Note that when
$\Delta_3$ is large enough to satisfy the condition $(\Omega_{\rm p}/\Delta_3)^2\ll
(\Omega_{\rm c}/\Gamma_3)^2$, the steady-state solution is reduced into the
simple form $\sigma_{11}^{(0)}\approx 1$,
$\sigma_{31}^{(0)}\approx \Omega_{\rm p}/d_{31}$, and all other
$\sigma_{jl}^{(0)}\approx 0$. This means that in the steady state  most atoms occupy
the ground state $|1\rangle$ but there are a small amount of atoms that occupy the excited state $|3\rangle$
will provide a gain to the signal field.

Note that when obtaining the above result on the steady-state solution we have assumed that there exist
finite dephasing rates $\gamma_{12}^{\rm col}$ and $\gamma_{14}^{\rm col}$ [see Eq.~(\ref{atomic1}) and Eq.~(\ref{atomic2})],
which make the populations in the metastable states $|2\rangle$ and $|4\rangle$ be zero (i.e. $\sigma_{22}^{(0)}=\sigma_{44}^{(0)}=0$) when $\Omega_p=0$.
If $\gamma_{12}^{\rm col}$ and $\gamma_{14}^{\rm col}$ are zero (they are usually small in realistic situation~\cite{Franzen}), before the opening of  $\Omega_p$ the populations in $|2\rangle$ and $|4\rangle$ may be comparable with that in the ground state $|1\rangle$. In this case, an initial preparation of the system is necessary (especially for the storage and retrieval of optical pulses of fast light discussed below), i.e., a technique such as optical pumping should be used to empty the populations in the metastable states $|2\rangle$ and $|4\rangle$
before $\Omega_p$ is opened.

When the weak signal field is switched on, the system will evolve into a time-dependent state. Solving the MB
Eqs.~(\ref{Blochequation}) and (\ref{Maxwell}), we obtain the solution with the form
$\Omega_{\rm s}\sim \exp \{i[K(\omega)z-\omega \tau]\}$, with the linear dispersion relation (for large $\Delta_3$) given by
\begin{equation}\label{Linear}
K(\omega)=\frac{\kappa_{23}\Omega_{\rm p}\sigma_{31}^{(0)}}{\Delta_3}\cdot
\frac{(\omega+d_{41})}{(\omega+d_{21})(\omega+d_{41})-|\Omega_{\rm c}|^2}.
\end{equation}

Shown in Fig.~\ref{fig1}(b) is the normalized gain spectrum of the signal
field, i.e. $-\text{Im}(K)/|\text{Im}(K)|_{\rm max}$, as a function
of $\omega$ and $\Omega_{\rm c}$. The system is chosen as an alkali
$^{85}$Rb atomic gas, with the levels $|1\rangle=|5^2S_{1/2},F=2,m_F=0\rangle$,
$|3\rangle=|5^2P_{1/2},F=2,m_F=1\rangle$,
$|2\rangle=|5^2S_{1/2},F=3,m_F=2\rangle$, and
$|4\rangle=|5^2S_{1/2},F=2,m_F=2\rangle$.
The system parameters are
$\Gamma_{3}=6~\text{MHz}$, $\gamma_{12}=10~\text{kHz}$,
$\gamma_{14}=10~\text{kHz}$, $\gamma_{24}=10~\text{kHz}$, $\Delta_3=2~\text{GHz}$,
$\Delta_2=\Delta_4=0$,
$\kappa_{23}=2\times10^{10}~\text{cm}^{-1}\text{s}^{-1}$, and
$\Omega_{\rm p}=1~\text{MHz}$. With these parameters we have $\sigma_{11}^{(0)}\approx0.999$, $\sigma_{22}^{(0)}\approx 6\times 10^{-4}$,
$\sigma_{33}^{(0)}\approx 1.0\times 10^{-6}$, and $\sigma_{44}^{(0)}\approx 6\times10^{-4}$.
From the figure we see that, when $\Omega_{\rm c}=0$, the gain spectrum is a single-peak profile
and has the maximum value at $\omega=0$ (blue solid line)~\cite{note01}. In this case the signal field grows rapidly during propagation.
When $\Omega_{\rm c}$ is increased to $0.1~\text{MHz}$  a {\it gain doublet} (i.e. two peaks with a minimum at the center frequency of the signal field) appears in the gain spectrum (red dashed line). This means that the gain of the signal field near center frequency
is largely suppressed by the introduction of the control field. When $\Omega_{\rm c}$ increases
further to $\Omega_{\rm c}=1~\text{MHz}$, the width and depth of the gain
doublet also increases (black dashed-dotted line). If a larger control field is used
(e.g. $\Omega_{\rm c}=10~\text{MHz}$), the width and depth of the gain
doublet will increase further. Consequently, the gain of the signal field
can be lowered to a very small value by the manipulation of the control field.
The appearance of the gain doublet in the system is due
to the quantum destructive interference effect induced by the control
field. A detailed theoretical analysis of the quantum interference effect in various ARG systems
have been given in Ref.~\cite{Xu}.

From the above result, we see that the system described in Fig.~{\ref{fig1}(a) has the following two attractive characteristics. (i)~The pump field coupling the quantum states $|1\rangle$ and $|3\rangle$ provides the gain to the signal field, and hence the signal field has no absorption during propagation. (ii)~The control field coupling the quantum states $|2\rangle$ and $|4\rangle$ contributes the quantum destructive interference effect that suppress the growth of the signal field (anti-gain). By suitably manipulating the pump and control fields, a balance between the gain and anti-gain can be achieved. In this way, the signal field can be very stable during propagation.

Shown in Fig.~\ref{fig2}
\begin{figure}
\includegraphics[scale=1.2]{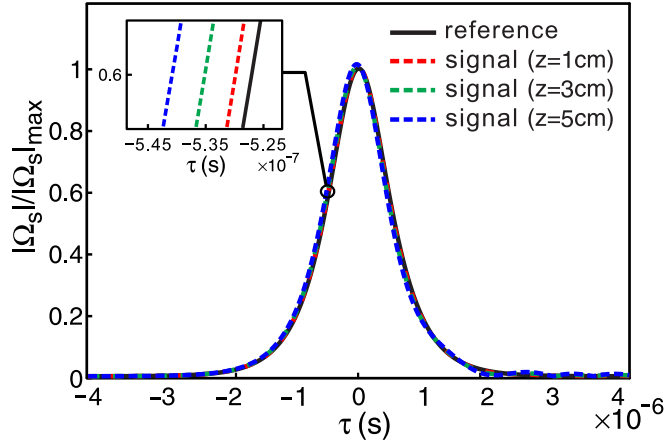}
\caption{(Color online) Stable propagation of fast light pulse via ARG. The dashed line is normalized signal pulse $|\Omega_{\rm s}|/|\Omega_{\rm s}|_{\rm max}$ as a function of $\tau$ when propagating to $z=1$ cm (red), $z=3$ cm (green), and $z=5$ cm (blue), respectively. The black line is a reference pulse that travels (in vacuum) with velocity $c$. The inset shows the advance of the signal pulse comparing with the reference pulse near $\tau=-5.25 \times 10^{-7}$ s. At $z=5$ cm (the blue dashed line) the signal pulse has a $14~\text{ns}$ advance comparing with the reference pulse.
}\label{fig2}
\end{figure}
%
is the numerical result of the signal pulse propagation by solving the MB Eqs.~(\ref{Blochequation}) and (\ref{Maxwell}) directly. When plotting the figure, the system parameters are chosen as $\Gamma_{3}=6~\text{MHz}$, $\gamma_{12}=10~\text{kHz}$, $\gamma_{14}=10~\text{kHz}$, $\gamma_{24}=10~\text{kHz}$, $\Delta_3=2~\text{GHz}$,
$\Delta_2=\Delta_4=0$, $\kappa_{23}=1\times10^{10}~\text{cm}^{-1}\text{s}^{-1}$, and $\Omega_{\rm p}=\Omega_{\rm c}=10~\text{MHz}$. The red, green, and blue dashed lines in the figure are for the normalized signal field  $|\Omega_{\rm s}|/|\Omega_{\rm s}|_{\rm max}$
as a function of $\tau$ propagating to $z=1$ cm (red), $z=3$ cm (green), and $z=5$ cm (blue), respectively. For comparison, a reference pulse
that travels in vacuum (i.e. with velocity $c$) is also shown (black solid line). The inset shows the amplification of all curves
near at $\tau=-5.25\times 10^{-7}$ s, the advance of the signal pulse comparing with the reference pulse can be clearly seen. At $z=5$ cm (the blue dashed line) the signal pulse has  an advance of $14~\text{ns}$ in comparison with the reference pulse. The result demonstrated in this figure indicates that the signal pulse is not only a fast light but also very stable, i.e. the pulse amplitude and pulse duration are nearly unchanged during propagation, which is desirable for practical applications, including the light memory using the present fast-light medium shown in the next subsection.

From the expression (\ref{Linear}) it is easy to calculate the value of the group velocity of the signal pulse by using the definition
$V_{g}=[\partial K/\partial \omega]^{-1}$, which can be either larger than $c$  or negative. In fact, using the system parameters given in Fig.~\ref{fig2} we obtain ${\rm Re}(V_{\rm g})=-1.33\times10^{-2}c$. Consequently, the signal light is a fast light and the system is a fast-light
medium.

We stress that the system used here is nearly free of Doppler effect and the spontaneous emission from the excited state. The reason is as follows. If the inhomogeneous broadening due to the atomic motion of center of mass is included, the linear dispersion relation for large $\Delta_3$ is given by
$K'(\omega)=[\kappa_{23}\Omega_{\rm p}\sigma_{31}/\Delta_3]\int^{\infty}_{-\infty} dv f(v) G(\omega,v)$. Here
$G(\omega,v)=(\omega+d_{41})/[(\omega+d_{21})(\omega+d_{41})-|\Omega_{\rm c}|^2]$,
$d_{41}=\Delta_4+i\gamma_{14}-(k_{\rm p}-k_{\rm c}-k_{\rm s})v$, $d_{21}=\Delta_2+i\gamma_{12}-(k_{\rm p}-k_{\rm s})v$, and $f(v)$ the Maxwell velocity distribution function satisfying $\int^{\infty}_{-\infty} dv f(v)=1$. Note that in our system the level structure allows $k_{\rm p}-k_{\rm c}-k_{\rm s}\approx 0$ and $k_{\rm p}\approx k_{\rm s}$, which mean $d_{41} \approx \Delta_4+i\gamma_{14}$ and $d_{21} \approx \Delta_2+i\gamma_{12}$. As a result, the function $G(\omega,v)$ has a negligible dependence on the atomic velocity $v$, and hence $K'(\omega)\approx K(\omega)$ (which is independent of $\gamma_{13}$), i.e. the Doppler effect and the spontaneous emission from the excited state $|3\rangle$ plays a negligible role in the system.

\subsection{Storage and retrieval of light pulses using the fast-light medium}{\label{Sec32}}

We now turn to explore the possibility for the realization of light memory in our ARG system. In some sense, light memory can be taken as an active manipulation of quantum states of atoms and light by using light. In an initial attempt, a numerical simulation is carried out based on the MB equations (\ref{Blochequation}) and (\ref{Maxwell}) for the storage and retrieval of signal pulse $\Omega_{\rm s}$ by always switching on the pump field $\Omega_{\rm p}$ but adiabatically switching-on and switching-off of the control field $\Omega_{\rm c}$ only. Similar to that suggested in Ref.~\cite{Fleischhauer2000} for EIT storage of slow light pulses, we firstly switch on the control field, then switch it off for some time interval, and at final stage we switch it on again. The result shows that in this way a storage of the signal pulse is not possible. The reason can be explained as follows. If the pump field $\Omega_{\rm p}$ is always switched on, the signal pulse has a large gain during the time interval in which the control field is switched off. Since $\Omega_{\rm c}$ is zero and thus no quantum destruction interference effect occurs, the signal pulse is amplified during the storage stage. Thus the signal pulse cannot be stored in the atomic medium by manipulating the control field only. Of course the retrieval of the signal pulse is also not possible because there is no storage occurs before the retrieval stage.

To overcome the difficulty stated above, one must find a way to avoid the amplification of the signal pulse during the storage stage. In a new attempt, we switch off also the pump filed $\Omega_{\rm p}$  when the control field $\Omega_{\rm c}$ is switched off. That is to say, the  pump and the control fields are designed to be adiabatically switched on and switched off {\it simultaneously} during the whole process of the storage and retrieval of the signal pulse. The simultaneous switching-on and switching-off of the pump and the control fields are modeled by the combination of two hyperbolic tangent functions with the form
\begin{subequations}\label{controlII}
\begin{eqnarray}
\Omega_{\rm p}(0,\tau)=\Omega_{\rm p0}\left\{1-\frac{1}{2}\tanh\left[\frac{\tau-T_{\text{off}}}{T_{\rm sp}}\right]
+\frac{1}{2}\tanh\left[\frac{\tau-T_{\text{on}}}{T_{\rm sp}}\right]\right\},\label{control1}\\
\Omega_{\rm c}(0,\tau)=\Omega_{\rm c0}\left\{1-\frac{1}{2}\tanh\left[\frac{\tau-T_{\text{off}}}{T_{\rm sc}}\right]
+\frac{1}{2}\tanh\left[\frac{\tau-T_{\text{on}}}{T_{\rm sc}}\right]\right\},\label{control2}
\end{eqnarray}
\end{subequations}
where $T_{\text{off}}$ and $T_{\text{on}}$ are respectively the times of switching-off and switching-on; $T_{\rm sp}$ and $T_{\rm sc}$ are the durations of the switching process respectively for the pump and control field; $\Omega_{\rm p0}$ and $\Omega_{\rm c0}$ are respectively the amplitudes of pump and control fields. The storage period (or called storage time) of the signal field is approximately given by $T_{\rm on}-T_{\rm off}$.
In the numerical simulation, the wave shape of the input signal pulse is taken to be a hyperbolic secant one, i.e., $\Omega_{\rm s}=\Omega_{\rm s0}\,\text{sech}(2\tau/\tau_0)$, where $\Omega_{\rm s0}$ is the characteristic amplitude of the signal pulse, which is taken to be much smaller than $\Omega_{\rm p0}$ and $\Omega_{\rm c0}$. System parameters are chosen to be $\Gamma_{3}=6~\text{MHz}$, $\gamma_{12}=10~\text{kHz}$, $\gamma_{14}=10~\text{kHz}$, $\Delta_3=1~\text{GHz}$, $\Delta_2=20~\text{MHz}$, $\Delta_4=-0.1~\text{MHz}$, $\kappa_{23}=1\times10^{10}~\text{cm}^{-1}\text{s}^{-1}$, $\Omega_{\rm s0}=100~\text{kHz}$, $\Omega_{\rm c0}=\Omega_{\rm p0}=10~\text{MHz}$, $\tau_0=1.0\times10^{-6}~\text{s}$, $T_{\rm sp}=1.5\times10^{-6}~\text{s}$, $T_{\rm sc}=0.5\times10^{-6}~\text{s}$, $T_{\text{off}}=2\times10^{-6}~\text{s}$, $T_{\text{on}}=15\times10^{-6}~\text{s}$, and $L=5~\text{cm}$ (medium length).

Shown in Fig.~\ref{fig3}(a)
\begin{figure}
\includegraphics[scale=0.75]{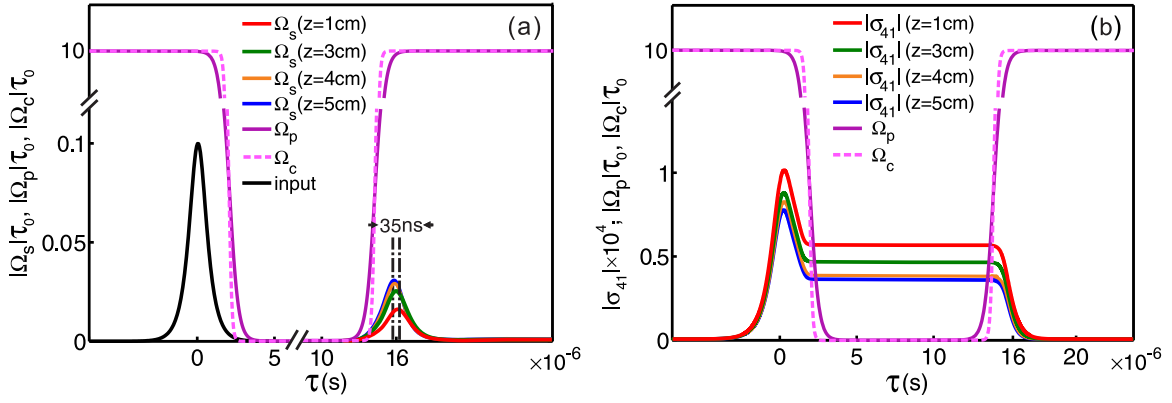}
\caption{(Color online) (a)~Storage and retrieval of the signal pulse using the fast-light medium via ARG. The purple solid and red dashed lines are respectively the pump ($|\Omega_{\rm p}|$) and the control ($|\Omega_{\rm c}|$) fields, which are switched on and off simultaneously. The black solid line is the input signal pulse ($|\Omega_{\rm s}|$) at the entrance ($z=0$) of the medium. The colored solid lines are the retrieved signal pulse at $z=1$ cm (red), $z=3$ cm (green), $z=4$ cm (yellow),  and $z=5$ cm (blue), respectively. The peak of the retrieved blue pulse has a advance of 35 ns comparing with the red pulse. (b)~Behavior of atomic coherence $|\sigma_{41}|$ during the light memory process.
Red, green, and blue solid lines are $|\sigma_{41}|$ as a function of $\tau$ at $z=1$ cm (red), $z=3$ cm (green), $z=4$ cm (yellow),  and $z=5$ cm (blue), respectively. The profiles of the pump and the control fields ($|\Omega_{\rm p}|$, $|\Omega_{\rm c}|$) are also
shown.}\label{fig3}
\end{figure}
%
is the result of the numerical simulation on the storage and retrieval of the signal pulse  $|\Omega_{\rm s}|$ via ARG.
The purple solid and red dashed lines in the figure are respectively the pump field $|\Omega_{\rm p}|$ and the control field $|\Omega_{\rm c}|$, both of them are switched on and switched off simultaneously. The black solid line is the input signal pulse at the entrance ($z=0$) of the medium. The colored solid lines denote the retrieved signal pulse at the position $z=1$ cm (red), $z=3$ cm (green), $z=4$ cm (yellow), and $z=5$ cm (blue), respectively. We find that: (i)~The signal pulse can be well stored and retrieved in the system; (ii)~The peak of the retrieved pulse at larger $z$ has an advance comparing with that at smaller $z$ . Especially the blue pulse has an advance of 35 ns comparing with the red pulse in the figure, which means that the retrieved pulse is indeed a fast light. Notice that the amplitude of the retrieved pulse is reduced in comparison with that of the input pulse. The reason is that an energy leakage of the fast light pulse cannot be avoided during the process of the storage and retrieval.

Although all the light fields become vanishing during the time interval of the storage, the excitation of atomic degrees, in particular atomic coherence, is not be zero. Fig.~\ref{fig3}(b) shows the atomic coherence $|\sigma_{41}|$ as a function of $\tau$ during the memory process of the signal pulse. The red, green, and blue solid lines are $|\sigma_{41}|$ at the positions $z=1$ cm (red), $z=3$ cm (green), $z=4$ cm (yellow), and $z=5$ cm (blue), respectively. The profiles of the pump and control fields ($|\Omega_{\rm p}|$, $|\Omega_{\rm c}|$) are also shown. We see that $|\sigma_{41}|$ has a non-zero value when both the pump and the control fields are switched off simultaneously. Thus the signal pulse is indeed stored and retained in the medium during the storage stage, which can be retrieved when both the pump and control fields are switched on again. Such phenomenon, similar to that occurs for the light memory via EIT-based slow-light media, may be understood as a transient atomic response to the switching-on and the switching-off of the pump and control fields~\cite{Akulshin2005,Lezama2006,Akulshin2010}.

To describe the quality of the light memory quantitatively, we define memory efficiency $\eta$ to be the energy ratio between the retrieved pulse and the input pulse, i.e. $\eta=\int^{+\infty}_{T_{\rm on}}|\Omega_{\text{out}}(\tau)|^2d\tau/\int^{T_{\rm off}}_{-\infty}|\Omega_{\text{in}}(\tau)|^2d\tau$.
With this formula, the memory efficiency of the result shown in Fig.~\ref{fig3}(a) for the retrieval at $z=5$ cm is found to be $32.6\%$. The reason of so low memory efficiency is caused by the energy leak of the fast light pulse, the decoherence and the limited propagation length of the system.

We can improve the memory efficiency of the signal pulse by manipulating the pump and control fields further. We assume the pump and control fields are still switched on and off simultaneously, but designed to make the signal field have no gain before the storage and have a small gain after the storage. Shown in Fig.~\ref{fig4}
\begin{figure}
\includegraphics[scale=0.75]{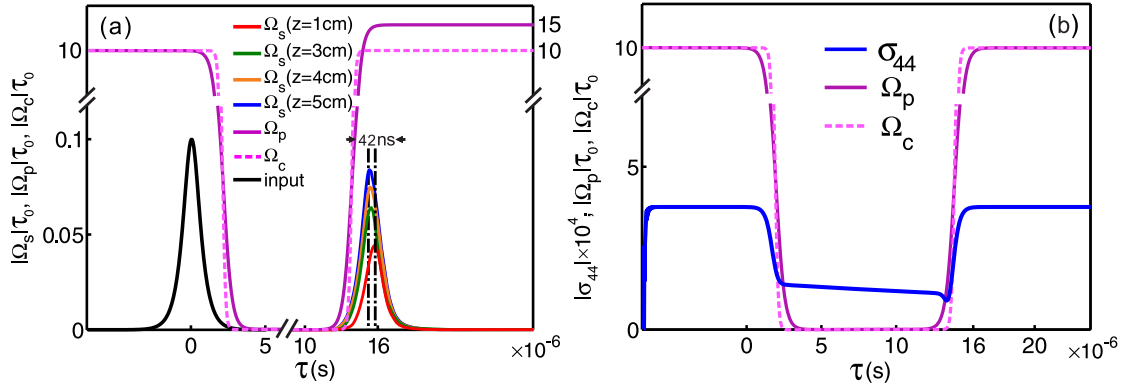}
\caption{(Color online) (a)~Improvement of the light memory in the fast-light medium. The pump field $|\Omega_{\rm p}|$ (purple solid line) and control field $|\Omega_{\rm c}|$ (purple dashed line) are still switched on and switched off simultaneously, but designed to make the signal field have no gain before the storage but have a small gain after the storage stage. The black solid line is the input signal pulse $|\Omega_{\rm s}|$ at the entrance ($z=0$). The colored solid lines are the retrieved signal pulse at $z=1$ cm (red), $z=3$ cm (green), $z=4$ cm (yellow),  and $z=5$ cm (blue), respectively. The peak of the retrieved blue pulse has an advance of 42 ns comparing with the red pulse. (b)~$\sigma_{44}$ during the memory process, as a function of $\tau$ (blue curve), which is amplified to $10^4$ times.}\label{fig4}
\end{figure}
is the result of the improved light memory. The amplitude $\Omega_{\rm c0}$ of the control field (purple dashed line) is designed to be the same before and after the storage, but the amplitude $\Omega_{\rm p0}$ of the pump field (purple dashed line) after the storage is designed to be larger than that before the storage. In this way, a small gain of the signal pulse after the storage is obtained to compensate for the energy leak of the fast light signal pulse during the storage stage. The black solid line in the figure is the input signal pulse $|\Omega_{\rm s}|$ at the entrance ($z=0$) of the medium. The colored solid lines are the retrieved signal pulse at $z=1$ cm (red), $z=3$ cm (green), $z=4$ cm (yellow),  and $z=5$ cm (blue), respectively. In this situation, the peak of the retrieved blue pulse has an advance of 42 ns comparing with the red pulse. When plotting the figure, the system parameters are chosen as $\Gamma_{3}=6~\text{MHz}$, $\gamma_{12}=10~\text{kHz}$, $\gamma_{14}=10~\text{kHz}$, $\Delta_3=1~\text{GHz}$, $\Delta_2=20~\text{MHz}$, $\Delta_4=-0.1~\text{MHz}$, $\kappa_{23}=1\times10^{10}~\text{cm}^{-1}\text{s}^{-1}$, $\Omega_{s0}=0.1~\text{MHz}$, $\Omega_{\rm c0}=10~\text{MHz}$, $\tau_0=1.0\times10^{-6}~\text{s}$, $T_{\rm sp}=1.5\times10^{-6}~\text{s}$, $T_{\rm sc}=0.5\times10^{-6}~\text{s}$, $T_{\text{off}}=0.4\times10^{-6}~\text{s}$, $T_{\text{on}}=15\times10^{-6}~\text{s}$. The amplitude of the pump field $\Omega_{\rm p0}$ is taken to be $10~\text{MHz}$  ($15~\text{MHz}$) before (after) the storage stage. The memory efficiency of the signal pulse for the retrieval at $z=5$ cm is increased into $\eta=90.6\%$.

\subsection{Theoretical explanation}{\label{Sec33}}

Now we give a simple explanation of the light memory via ARG presented in the last subsection. From Fig.~\ref{fig1}(a) we note that the strong pump field $\Omega_{\rm p}$  coherently pumps the atoms in the system from the ground state $|1\rangle$ to the excited state $|3\rangle$, which provides a gain to the signal field $\Omega_{\rm s}$ (when $\Omega_{\rm s}$  is switched on). That is to say, the role of the ground state $|1\rangle$ and the pump field $\Omega_{\rm p}$ is to provide an optical pumping that initially prepares a population in the excited state $|3\rangle$.
When the one-photon detuning $\Delta_3$ is taken to be large enough ($\sim$ 1 GHz) for suppressing the gain and Doppler effect of $\Omega_{\rm s}$, the levels $|3\rangle$, $|2\rangle$, $|4\rangle$, $\Omega_{\rm s}$, and the control field $\Omega_{\rm c}$ constitute a three-level ladder system with a small population in the state $|3\rangle$.

The role of the control field $\Omega_{\rm c}$ is to bring a quantum interference effect to suppress the gain of the signal field $\Omega_{\rm s}$. Due to the introduction of the control field, the states $|2\rangle$ and $|4\rangle$ mix together. As a result, two dressed states form, which leads to a destructive interference to the signal-field amplitude and thereafter the occurrence of the doublet of the gain spectrum of the signal field, as illustrated in Fig.~\ref{fig1}(b).
It is just the quantum destructive interference effect induced by the control field $\Omega_{\rm c}$ in the three-level ladder systems that make the stable propagation and memory of the fast-light signal pulses possible.

The light memory of the present scheme can also be understood based on the solution of the MB Eqs.~(\ref{Blochequation}) and (\ref{Maxwell}).
Using the steady-state solution given in Sec.~\ref{Sec31} and  assuming $\Delta_3\gg\gamma_{ij}$, $\Gamma_{ij}$, $\Omega_{\rm p}$, $\Omega_{\rm c}$,
Eqs.~(\ref{Blochequation}) and (\ref{Maxwell}) under the linear approximation can be simplified into
\begin{subequations}\label{3Eq}
\begin{eqnarray}
&&i\frac{\partial}{\partial
t}\sigma_{21}+d'_{21}\sigma_{21}+\Omega_{\rm c}\sigma_{41}+\Omega^*_{\rm s}\sigma^{(0)}_{31}=0,\\
&&i\frac{\partial}{\partial t}\sigma_{41}+\Omega_{\rm c}\sigma_{21}=0,\\
&&i\left(\frac{\partial}{\partial z}+\frac{1}{c}\,\frac{\partial}{\partial t}\right)\Omega_{\rm s}-\frac{\kappa_{23}\Omega_{\rm p}}{\Delta_3}\sigma^*_{21}=0,
\end{eqnarray}
\end{subequations}
with $d'_{21}=d_{21}+\Omega^2_{\rm p}/\Delta_3$ and $\sigma^{(0)}_{31}=-\Omega_{\rm p}/\Delta_3$. When deriving the above result, we have returned to original $z$-$t$ coordinates and assumed that $\Omega_{\rm c}$ and $\Omega_{\rm p}$ are real and vary slowly with time $t$. From Eq.~(\ref{3Eq}) we obtain $\sigma_{21}=-(i/\Omega_{\rm c}) \partial \sigma_{41}/\partial t$, and
\begin{subequations}\label{2slv}
\begin{eqnarray}
\sigma_{41}& & =-\frac{\Omega^*_{\rm s}}{\Omega_{\rm c}} \sigma^{(0)}_{31}+
                \frac{i}{\Omega_{\rm c}}\left[\left(\frac{d'_{21}}{\Omega_{\rm c}}+\frac{\partial}{\partial t}\frac{i}{\Omega_{\rm c}}\right)\frac{\partial}{\partial t}\sigma_{41}\right], \nonumber\\
           & & \approx -\frac{\Omega^*_{\rm s}}{\Omega_{\rm c}} \sigma^{(0)}_{31}=\frac{\Omega^*_{\rm s}\Omega_{\rm p}}{\Delta_3 \Omega_{\rm c}}.
\end{eqnarray}
\end{subequations}
This result tells us that, although the pump, signal, and control fields are decreased to zero simultaneously during the storage stage, the coherence of the atoms, $\sigma_{41}$, can maintain to be a finite constant value if the changing rate of the product of the signal and pump fields (i.e. $\Omega_{\rm s}^* \Omega_{\rm p}$) and the changing rate of the control field ($\Omega_{\rm c}$) are the same. This is just the case considered in our numerical simulation for the light memory obtained in Fig.~3 and Fig.~4.

\section{Discussion and summary}{\label{Sec4}}

Finally, we make some remarks on the calculation presented above. First, in our calculation a semi-classical approach is employed, which is valid since the photon number $N$ in the signal pulse  has been taken to be large ($N\approx 10^5\sim 10^6$). Second, the noise due to spontaneous emission  is negligible in our system. The reasons are the following: (i)~The population in the excited state $|3\rangle$ is quite small ($\sigma_{33}\approx 10^{-5}\sim10^{-6}$), which is due to the large one-photon detuning $\Delta_3$ ($\approx$ 1 GHz) we have chosen; (ii)~Although the system is a gain one, the gain acquired by the signal pulse is vanishing small because it is suppressed largely by the quantum destructive interference effect resulted from the strong control field. Third, because there is no significant population in the state $|4\rangle$, the coupling between the states $|3\rangle$ and $|4\rangle$ by the pump field is negligible, and hence the noise induced by four-wave mixing effect plays a negligible role in our system. To prove the non-significant population in $|4\rangle$, we have calculated $\sigma_{44}$ in the system, with the result plotted in Fig.~\ref{fig4}(b). We see that $\sigma_{44}$ is of the order of magnitude $10^{-4}$  before, during, and after the storage process, which is indeed very small and thus can be neglected.

Note that in an interesting work Agarwal {\it et al.}~\cite{Aga2} showed that a three-level $\Lambda$ system with a microwave field coupling two lower states (called closed $\Lambda$ system) can support a transition of pulse propagation from subluminal to superluminal. In the following years, much attention has been paid to theoretical and experimental studies on closed $\Lambda$ systems (see, e.g. Refs.~\cite{Hebin,Kor,Man}). However, the closed $\Lambda$ systems used in all those works~\cite{Aga2,Hebin,Kor,Man} are very different from our system shown in Fig.~\ref{fig1}(a). First, our system is not closed and thus no dependence of the relative phase between the signal, pump, and control fields. Second, in Refs.\cite{Aga2,Hebin,Kor,Man} no study was carried out for the storage and retrieval of fast light. Third, in the ARG scheme a stable propagation of fast light is not possible if working with a closed $\Lambda$-type level configuration. Although in our present work only the memory of classical light pulses is considered, the approach can be generalized to the case of memory for quantized light pulses. The advantages of our ARG scheme, such as very weak four-wave mixing noise and immunization of Doppler effect are promising for quantum memory.

In conclusion, in this work we have proposed a scheme for realizing a light memory in a four-level atomic system via an ARG mechanism. We have shown that a stable propagation of signal light pulses with superluminal velocity is possible through the ARG contributed by the pump field and the quantum destructive interference effect induced by the control field. We have demonstrated that a robust storage and retrieval of the signal pulses can be achieved by switching on and off the pump and the control fields simultaneously. The research and results reported here opened up a new way of light memory, and is promising for practical applications of light information processing and transmission using fast-light media.


\acknowledgments The authors would like to thank C. Zhu and C. Hang for fruitful discussions. This work was supported by NSF-China under Grants No.~11475063 and No.~11474099.

\appendix

\section{Expression of the optical Bloch equation}\label{ap1}

The explicit expressions of the optical Bloch equation reads
\begin{subequations}\label{atomic1}
\begin{eqnarray}
&&i\frac{\partial}{\partial
\tau}\sigma_{11}-i\Gamma_{12}\sigma_{22}-i\Gamma_{13}\sigma_{33}-i\Gamma_{14}\sigma_{44}+
\Omega^*_{\rm p}\sigma_{31}-\Omega_{\rm p}\sigma^*_{31}=0,\\
&&i \frac{\partial}{\partial
\tau}\sigma_{22}-i\Gamma_{23}\sigma_{33}+
\Omega^*_{\rm s}\sigma_{32}-\Omega_{\rm s}\sigma^*_{32}+\Omega^*_{\rm c}\sigma_{42}-\Omega_{\rm c}\sigma^*_{42}=0,\\
&&i\left(\frac{\partial}{\partial
\tau}+\Gamma_3\right)\sigma_{33}+\Omega_{\rm s}\sigma^*_{32}+\Omega_{\rm p}\sigma^*_{31}-\Omega^*_{\rm s}\sigma_{32}-\Omega^*_{\rm p}\sigma_{31}=0,\\
&&i\frac{\partial}{\partial
\tau}\sigma_{44}-i\Gamma_{42}\sigma_{22}+\Omega_{\rm c}\sigma^*_{42}+\Omega^*_{\rm c}\sigma_{42}=0,
\end{eqnarray}
\end{subequations}
for diagonal matrix elements, and
\begin{subequations}\label{atomic2}
\begin{eqnarray}
&&\left(i\frac{\partial}{\partial
\tau}+d_{21}\right)\sigma_{21}+\Omega^*_{\rm c}\sigma_{41}+\Omega^*_{\rm s}\sigma_{31}-\Omega_{\rm p}\sigma^*_{32}=0,\label{A2a}\\
&&\left(i\frac{\partial}{\partial
\tau}+d_{31}\right)\sigma_{31}+\Omega_{\rm p}(\sigma_{11}-\sigma_{33})+\Omega_{\rm s}\sigma_{21}=0,\\
&&\left(i\frac{\partial}{\partial
\tau}+d_{32}\right)\sigma_{32}+\Omega_{\rm p}\sigma^*_{21}+\Omega_{\rm s}(\sigma_{22}-\sigma_{33})-\Omega_{\rm c}\sigma^*_{43}=0,\label{A2c}\\
&&\left(i\frac{\partial}{\partial
\tau}+d_{41}\right)\sigma_{41}+\Omega_{\rm c}\sigma_{21}-\Omega_{\rm p}\sigma_{43}=0,\label{A2d}\\
&&\left(i\frac{\partial}{\partial
\tau}+d_{42}\right)\sigma_{42}+\Omega_{\rm c}(\sigma_{22}-\sigma_{44})-\Omega_{\rm s}\sigma_{43}=0,\\
&&\left(i\frac{\partial}{\partial
\tau}+d_{43}\right)\sigma_{43}+\Omega_{\rm c}\sigma^*_{32}-\Omega^*_{\rm p}\sigma_{41}-\Omega^*_{\rm s}\sigma_{42}=0,\label{A2f}
\end{eqnarray}
\end{subequations}
for non-diagonal matrix elements. Here $d_{21}=\Delta_2+i\gamma_{21}$,
$d_{31}=\Delta_3+i\gamma_{13}$, $d_{32}=\Delta_3-\Delta_2+i\gamma_{23}$,
$d_{41}=\Delta_4+i\gamma_{14}$, $d_{42}=\Delta_4-\Delta_2+i\gamma_{24}$,
$d_{43}=\Delta_4-\Delta_3+i\gamma_{34}$; $\Delta_3=\omega_{\rm p}-(\omega_3-\omega_1)$,
$\Delta_2=\omega_{\rm p}-\omega_{\rm s}-(\omega_2-\omega_1)$, and
$\Delta_4=\omega_{\rm p}-\omega_{\rm s}-\omega_{\rm c}-(\omega_4-\omega_1)$
are respectively the one-, two-, and three-photon detunings, with $\hbar \omega_j$ the
eigenenergy of the state $|j\rangle$ ($j=1,2,3,4$);
$\gamma_{ij}=(\Gamma_{i}+\Gamma_{j})/2+\gamma^{\text{col}}_{ij}$, $\Gamma_j=\sum_{i<
j}\Gamma_{ij}$, with $\Gamma_{ij}$ the spontaneous emission decay rate and
$\gamma^{\text{col}}_{ij}$ the dephasing rate between state $|i\rangle$ and state
$|j\rangle$~\cite{Boyd}.



\begin{thebibliography}{}
\bibitem{note00}Slow light (fast light) refers to the situation in which the group
velocity $V_g$ of the light pulse satisfies $V_g<<c$\,  ($V_g>c$ or negative).
\bibitem{note01}The frequency and wave number of the signal field are given by
$\omega_{\rm s}+\omega$ and $k_{\rm s}+K(\omega)+\omega/c$, respectively. Thus $\omega=0$ corresponds to the central frequency
(i.e., $\omega_{\rm s}$) of the signal field.
\end{thebibliography}
\end{document}